\documentstyle[psfig,draft]{mn}

\input epsf

\begin{document}

\title[The  neutrino background]
{Constraints on the cosmic neutrino background }
\author[E.~Pierpaoli] {Elena Pierpaoli${}^1$\\
${}^1$Physics Department and Astronomy Department,
 Princeton University, Princeton, NJ, 08544~~USA \\}

\date{Accepted ... ;
      Received ... ;
      in original form ...}

\pagerange{000--000}

\maketitle

\begin{abstract}
The radiative component of the Universe has a characteristic impact on 
both large scale structure  (LSS) and the cosmic microwave background radiation (CMB).
We use the recent WMAP data, together with previous CBI data and 2dF matter
power spectrum, to constrain the effective number of neutrino species 
$N_{eff}$ in a general Cosmology.
We find that $N_{eff}=4.31$ with a 95 per cent C.L. $1.6 \le  N_{eff} \le   7.1$.
If we include the $H_0$ prior from  the HST project
we find the best fit $N_{eff}=4.08$ 
 and   $1.90 \le N_{eff} \le 6.62$ for  95 per cent C.L.
The curvature we derive is still consistent with flat, but
assuming a flat Universe  from the beginning 
implies a bias toward lower $N_{eff}$, as well as 
artificially smaller error bars. Adding the Supernovae constraint
doesn't improve the result. 
We analyze and discuss the degeneracies with other parameters, and point
out that probes of the matter power spectrum on smaller scales, accurate
independent $\sigma_8$ measurements, together
with better independent measurement of $H_0$ would help in breaking the 
degeneracies. 

\end{abstract}


\section{Introduction \label{sec:intro}}
The quest for cosmological parameters is the major goal of cosmology, and we
are experiencing an exciting era in which newer and better data
allow for the precise determination of an increased number of parameters.
In this paper we focus on one specific parameter,
the amount of relativistic energy density at recombination/equivalence, 
and we discuss the implications of the  current CMB and LSS observations.
The relativistic energy 
density  is often parametrized by the equivalent number of standard
model neutrino species  $N_{eff}$.
The standard model predicts three neutrino species, plus a  correction 
 imposed by neutrinos not being completely decoupled during 
electron-positron annihilation
 ($\Delta N_{eff}=0.03$) (Dolgov et al. 1997, Gnedin \& Gnedin 1998) 
 and by the finite temperature QED correction 
to the electromagnetic plasma ($\Delta N_{eff}=0.01$) (Lopez et al. 1999, 
Mangano et al. 2002).

Departures from the standard model which imply a variation in $N_{eff}$
may be due to  decaying dark matter (Bonometto \& Pierpaoli 1998, 
Hannestad 1998, Lopez et al 1998, Kaplinghat \& Turner 2001 
 or quintessence (Bean et al 2001).
In these  scenarios $N_{eff}$ can be either smaller or greater
than 3.04, and the constraints on  $N_{eff}$
 implied by 
CMB and LSS  need not to agree with the ones implied by Big Bang Nucleosyntesis
(Kaplinghat \& Turner 2001).
We will  analyze and discuss the CMB+LSS results on $N_{eff}$.
Unlike previous work on this subject, we allow for a  cosmology with 
general curvature. 
We will use the CMB data from the WMAP satellite (Bennet et al. 2003), 
and complement 
them with the CBI ones (Pearson et al. 2002).
As for the LSS, we use the 2dF data (Percival et al 2002).
We use a general cosmology in our analysis because at the present time 
we do not have 
an independent probe of the flatness of the Universe other than the CMB.
It is therefore fictitious and misleading to restrict ourselves to flat
cosmologies in the so called ``precision cosmology'' era.
We will show that allowing for non--zero curvature changes quite 
significantly  the results on  $N_{eff}$, despite the fact that the inferred 
 constraints on curvature are still compatible with a flat Universe.

The paper is organized as follows:
in section \ref{sec:effects} we discuss the effects of $N_{eff}$ on the
 matter and radiation power spectra, in section  \ref{sec:cmbcon}
we discuss the CMB+LSS  constraints,
 while section  (\ref{sec:con}) is dedicated to conclusions.

\section{ Effects of a relativistic background on the matter and radiation
power spectra}\label{sec:effects}

An excess of energy in relativistic particles with respect to the 
one predicted by the standard model has an impact on both the radiation
and the matter power spectrum, for different reasons.
The increased relativistic energy 
delays equivalence, and  scale entering the horizon 
 at equivalence is bigger.
   Since the amplitude of  dark matter  perturbations 
on smaller scales is frozen during the radiation dominated era,
the matter power spectrum turnover is shifted toward lower $k$'s.
The effect can be quantified via the shape parameter 
$\Gamma \simeq \Omega_m h (g_\star/3.36)^{1/2}$ (White, Gelmini \& Silk 1995), 
where $g_\star$
represents the relativistic degrees of freedom  ($g_\star= 2+0.454N_{eff}$).
For a given power at large scales, the small scale power s therefore reduced, 
and a lower value of $\sigma_8$ is implied.

An increased radiation density  implies a  
smaller conformal time
and therefore a smaller sound horizon. 
As a consequence, the peaks of the CMB power spectrum, which reflects
the peaks/valleys of the oscillations at recombination, 
are shifted toward smaller scales (bigger $l$'s) (Pierpaoli \& Bnometto 1999).
In addition, since
 recombination occurs when the Universe is not completely matter 
dominated, the early Integrated Sachs-Wolfe  (ISW) effect causes  an enhanced  first peak.

A description of parameter degeneracies  in a flat Universe
is given by Bowen et al. (2002), where they find that $N_{eff}$ 
is mainly  degenerate with $\Omega_m$ and the spectral index $n$.
They show that almost complete degeneracy can be obtained by
keeping $z_{eq}$, $\Omega_b h^2$ and $R$ (the position of the
 first acoustic peak with respect to a reference model) 
fixed while varying $N_{eff}$ produces  almost degenerate power spectra.
Here we extend the analysis to the real data and consider general curvature.

Allowing for possible curvature  affects the radiation power 
spectrum. In particular, the position of the peaks is strongly dependent on 
curvature. 
A given comoving scale at the last scattering surface subtends  an angle
on the sky that is curvature--dependent. If the Universe is open ($\Omega_k 
> 0 $) [ or closed ($\Omega_k < 0$)] then the angle associated with a 
particular scale at last scattering is smaller [larger] than
the corresponding flat Universe one. Consequently, 
the peaks/valleys of the CMB power spectrum 
will be located at larger [smaller] $l$'s.
A slightly close  Universe can compensate  the shift 
of the peaks toward higher $l$'s implied by 
an increased radiation background.
We therefore expect some  degeneracy between curvature and $N_{eff}$, and if
we aim to determine both from the CMB measurements, we ought to treat 
both of them  as free parameters in the data analysis.

\section{ CMB and LSS constraints}\label{sec:cmbcon}

We used  a modified version of the COSMOMC (Lewis \& Bridle 2002)
  package to compute the  likelihoods also including the WMAP new results (Verde et al. 2003).
We considered only CDM adiabatic perturbations, and the following set of 
parameters, with usual definitions:
   $\Omega_{dm}$,  $\Omega_\Lambda$, $\Omega_b$,
 $\Omega_k$, $H_0$, $N_\nu$, $n_s$, $A_s$, $z_{re}$ (reionization redshift).
In our analysis we use the  WMAP and CBI data together 
with the 2dF power spectrum, and we discuss the effect of adding 
other priors.

\begin{figure}
\begin{center}
\leavevmode
\epsfxsize=8cm \epsfbox{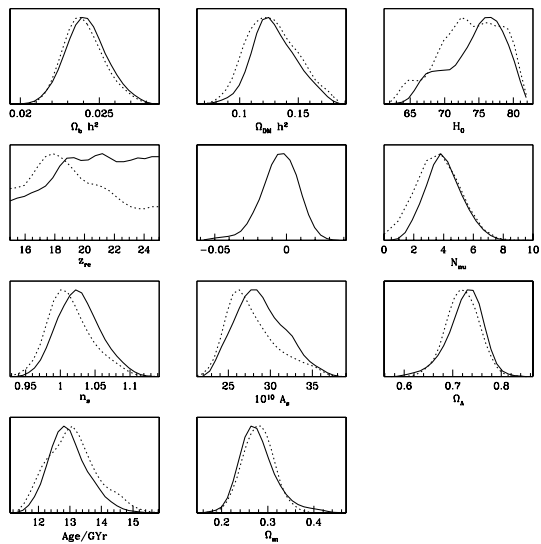}
\end{center}
\caption{The marginalized likelihoods for the parameters under consideration.
We have assumed here a top--hat prior on $H_0$ corresponding to the 1$\sigma$
interval allowed by the HST key project results. 
The solid line is  for a general curvature, the dotted corresponds to the flat Universe case. The general curvature tends to push the constraints on $N_{eff}$
toward higher values. The same upper limits on $N_{eff}$ are probably due to the upper limit imposed on $H_0$ ($<80$). 
 Notice that $z_{re}$ in the range considered here is not constrained by the
 data. }
\label{fig:1dlik}
\end{figure}

First we compare the results for a flat model ($\Omega_k=0$) with those for 
a general Universe when $H_0$ to vary only in the range $64 \le H_0 \le 80$ corresponding to the 1$\sigma$ allowed range of the HST result (Freedman et al. 2001).
In fig.\ref{fig:1dlik} we present the  likelihoods for each parameter, after 
marginalization over all the others, for flat 
and generally curved cases.  
For the number of effective neutrinos we find a best fit $N_{eff}=4.08$
($N_{eff}=3.70$) for the general  curvature (flat) case,
with   $2.23 \le N_{eff} \le 6.13$  ($1.82 \le N_{eff} \le 5.74$)
 at the 95\% C.L.
The assumption of a flat  Universe biases the result toward low $N_{eff}$
and shrinks the inferred error bars. 
Our results are in agreement with what found by Crotty et al. (2003) with the
assumption of flat geometry.
Notice that the general curvature analysis is important at the present time
because of the great improvement that the WMAP results have implied on
the CMB power spectrum.
Constraints from CMB and LSS on flat Universes  previous to the  WMAP
 were much weaker: $N_{eff} \le 15$ at 95\% C.L.
(Hannestad 2001, see also Hansen et al., 2002).

In fig.\ref{fig:1dlik} the flat and curve case present the similar 
upper limit for $N_{eff}$ mainly because we 
imposed a tight top-hat prior on $H_0$ ($<80$).
 The  likelihood  indeed seems to prefer high values for  $H_0$, 
which may allow for extra radiative component.

For the general curvature case, we now proceed to discuss the effect of
various priors on the determination of $N_{eff}$.

\begin{figure}
\begin{center}
\leavevmode
\epsfxsize=8cm \epsfbox{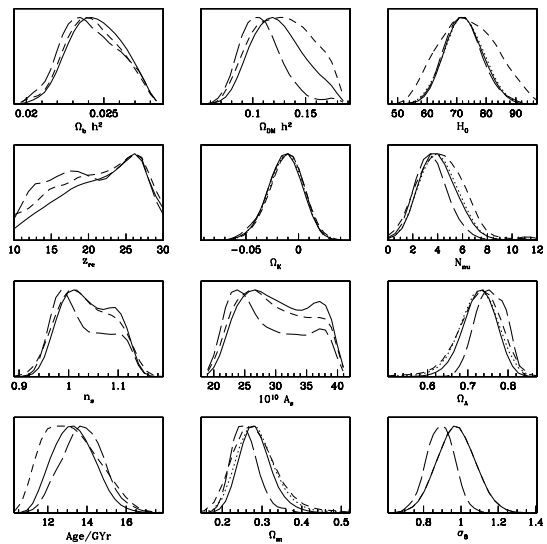}
\end{center}
\caption{The marginalized likelihoods in the case of a general cosmology.
The short--dashed line only consider CMB+2dF data, the dotted includes the $H_0$ prior
from the HST project, and the solid also includes SN data. 
$N_{eff}$ is restricted to be $\le 6.6$ at 95 per cent C.L.,
 and $\Omega_k$ tends to be negative 
but is still consistent with flat.
Note the high $n_s$ and $z_{re}$ values.
The long--dashed line is obtained adding an hypothetical prior on $\sigma_8$
with the typical scaling from clusters and weak lensing.
  }
\label{fig:f2}
\end{figure}

In fig.~\ref{fig:f2} we present the results  obtained allowing for a wider 
$H_0$ (and $z_{re}$) range.
The best fit for $N_{eff}$ in this case is 
$N_{eff}=4.31$,
 with a 95 per cent
C.L. range : $1.6 \le N_{eff} \le 7.1$ from CMB and 2dF only, and 
$N_{eff}=4.08$ with  $1.9 \le  N_{eff} \le   6.62$
when the $H_0$ prior from the HST project is included.

As for the other parameters, we note that the inclusion of extra relativistic 
energy causes a higher $n_s$ values than in the standard case 
(Spergel et al 2003), a higher $z_{re}$ and a slightly closed Universe
 ($\Omega_k=-0.013 \pm 0.015$). We argue that, since a high $N_{eff}$
 boosts the first peak through 
the early ISW  effect, a higher $n_s$ combined with 
an early reionization of the Universe can still ensure a good fit to the
 CMB data. 
Matter power spectrum data on smaller scales than the ones probed by 
2dF (e.g. Ly-$\alpha$ forest data) may be used in breaking this degeneracy,
because the matter power spectrum is only sensitive to $n_s$ and not 
to $z_{re}$. Data of the matter power spectrum at very small scales would
be incredibly sensitive to small increases in $n_s$.
However, there is still much debate on small scale data, since their 
interpretation  may be complicated by the non--linear growth of the
 fluctuations. 
We adopted here a conservative approach and chose not to include them 
in the analysis.
In fig.~\ref{fig:f2} we note that the derived value of $\sigma_8$ is quite 
high ($0.97 \pm 0.1$) yet because of the favoured high $n_s$  values.
Actual quotations of the $\sigma_8$ value range between 0.75 and 1 ( Pierpaoli et al. 2003, and references therein), so that
a broad prior on $\sigma_8$ would already decrease the allowed $n_s$ 
and improve the errors on $N_{eff}$.
Consistent  5 per cent accuracy measurements of $\sigma_8$  from 
both weak lensing and clusters 
would greatly improve the constraint on $N_{eff}$.
As an example, in fig.~\ref{fig:f2} we plot the curves 
obtained with an hypothetical 
prior on $\sigma_8$ with the typical scaling derived from
 cluster and weak lensing: 
$\sigma_8 ~ (\Omega_m/0.3)^{0.6} = 0.85 \pm 0.05$. 
Such prior would imply a lower $\Omega_m$ and $N_{eff}=3.8 \pm 1.6$.

As for the other parameters, we note that $\Omega_m$ and $\Omega_{\Lambda}$
are well constrained in the  range: 
$\Omega_m = 0.29 \pm 0.06$ and $\Omega_\Lambda = 0.72 \pm 0.05$ 
(1 $\sigma$ error).

Hannestad (2003) while analyzing flat Universes uses 
a tight prior on $\Omega_m$ derived from the Supernova Cosmology Project (Perlmutter e al. 1999).
As a consequence he derives a small $N_{eff}$ value  with  small error bars.
We run a separate chain treating CMB+LSS+SN data in a general cosmology with
COSMOMC,  and found that the inclusion of the Supernovae in the analysis 
doesn't particularly improve the results on $N_{eff}$ (see fig.~\ref{fig:f2}).
Including the SN in the analysis  slightly  improves the $\Omega_m$ and 
$\Omega_{\Lambda}$ determination, but the errors on $N_{eff}$ are practically 
unchanged.
We conclude that the strong constraints  obtained by  Hannestad (2003) are 
mainly due to the restriction to a flat Cosmology and to the artificially 
small prior assumed on $\Omega_m$ as a way of introducing the SN constraint.

In table \ref{tab:corrcur}
 we plot the correlation matrix for the parameters under 
consideration.
Note that $N_\nu$ is mostly correlated with $\Omega_{dm}h^2$ and $\Omega_k$. 
A better independent estimate of the redshift of equivalence 
(typically probing $\Omega_m h$), combined with 
a better independent measurement of $H_0$, may help in breaking the degeneracy.

\begin{table*}
\caption{\footnotesize%
 Correlation matrix for the parameters in fig~\ref{fig:1dlik}. We assume here 
a general curvature, and $64 \le H_0 \le 80$.}

\label{tab:corrcur}
\begin{center}
\begin{tabular}{l|ccccccccccc}
--&$\Omega_bh^2$ & $\Omega_{dm}h^2$ & $H_0$ & $z_{re}$ & $\Omega_k$ & $N_\nu$ & 
$n_s$ & $A_s$ & $\Omega_\Lambda$ & Age & $\Omega_m$\\
\hline
 $\Omega_bh^2$ &  1.00 & -0.29  & 0.14 &  0.39  & 0.15 & -0.44 &  0.78 &  0.70 &  0.28 &  0.07 & -0.30 \\
$\Omega_{dm}h^2$ &  -0.29  & 1.00  & 0.25 & -0.07 & -0.10 &  0.76 & -0.08 & -0.06 & -0.69 & -0.81  & 0.65\\
 $H_0$ &   0.14 &  0.25 &  1.00 &  0.11 &  0.48 &  0.17 &  0.22 &  0.16  & 0.46  & -0.75 & -0.57\\
 $z_{re}$ &  0.39 &  -0.07  & 0.11 &  1.00 & -0.38 &  0.18  & 0.71  & 0.87 &  0.27 &  0.01 & -0.12\\
 $\Omega_k$ &  0.15 & -0.10 &  0.48 & -0.38 &  1.00 & -0.58 & -0.16 & -0.10 &  0.17  &-0.28&  -0.47\\
$N_\nu$ &  -0.44 &  0.76 &  0.17 &  0.18  &-0.58 &  1.00 & -0.05&  -0.08&  -0.35 & -0.56 &  0.50\\
$n_s$ &   0.78 & -0.08 &  0.22 &  0.71  &-0.16 & -0.05 & 1.00  & 0.89 &  0.28&  -0.08 & -0.20\\
 $A_s$ &  0.70  &-0.06  & 0.16  & 0.87 & -0.10 & -0.08 &  0.89 &  1.00 &  0.20  &-0.05 & -0.14\\
 $\Omega_\Lambda$ &  0.28 & -0.69 &  0.46  & 0.27 &  0.17 & -0.35  & 0.28 &  0.20  & 1.00 &  0.22 & -0.95\\
 Age&  0.07 & -0.81 & -0.75 &  0.01 & -0.28 & -0.56&  -0.08&  -0.05 &  0.22 &  1.00 & -0.11\\
 $\Omega_m$& -0.30 &  0.65 & -0.57 & -0.12 & -0.47 &  0.50 & -0.20&  -0.14 & -0.95  &-0.11 &  1.00\\

\hline
\end{tabular}
\end{center}
\end{table*}

\section{Conclusions}\label{sec:con}
The new CMB data  (WMAP and CBI)  together with the matter power spectrum derived by the 2dF galaxy survey  can constrain the effective number of 
neutrino species much more precisely than previous experiments.
Previous estimates of $N_{eff}$ were derived under the assumption of nul 
curvature.
 Since we don't have any independent
 confirmation of the flatness of the Universe other than the CMB itself,
 we argue that
the curvature should be kept as a free parameter in the estimation of 
$N_{eff}$. We compare the results derived from the two different hypothesis.
Applying a top hat prior for $H_0$  ($64 \le H_0 \le 80$), 
we find for a flat Universe  $1.82 \le N_{eff} \le 5.74$ at 95 per cent C.L., with a best fit of $N_{eff}=3.70$, while with general curvature $2.23 \le N_{eff} \le 6.13$  with a best fit $N_{eff}=4.08$ .
Allowing for general curvature  shifts the acceptance range 
of $N_{eff}$ toward higher values mainly because the curvature tends to
compensate the effect of $N_{eff}$ on the peak locations in the CMB
 power spectrum.

In the case of general curvature, we  explored a wider range in 
$H_0$ and $z_{re}$ and applied different priors.
We find  $1.6 \le N_{eff} \le 7.1$ (best fit $N_{eff}=4.31$)
 at 95\% C.L.   from CMB and 2dF only, and 
$N_{eff}=4.08$ with  $1.9 \le  N_{eff} \le   6.62$
when the $H_0$ prior from the HST project is included as a proper Gaussian
 prior.
No significant modifications derive from the inclusion of the SN constraint.

By looking at the likelihood distribution for each parameter after 
marginalization over all the others we conclude that the inclusion 
of an extra relativistic component would  suggest a higher 
expansion rate $H_0$, a higher spectral index $n_s$ and  
$\Omega_k$ slightly negative, if compared to the standard 
analysis with three neutrinos (Spergel et al. 2003).

We analyze the correlations between the various parameters and conclude that
$N_{eff}$ is most degenerate with $\Omega_{dm}h^2$ and $\Omega_k$.
We argue that other independent measurements of the matter power spectrum,
like precise determinations of  $\sigma_8$ from clusters and lensing or probes
 at smaller scales from the Ly--$\alpha$ forest,  would help in constraining the 
epoch of equivalence and therefore would improve the results on $N_{eff}$.
 Moreover, it would greatly improve 
the constraints on the large $n_s$ now allowed.

\section{Acknowledgments}
EP is supported by NASA grant NAG5-11489.
The author thanks Antony Lewis for assistance with the use of COSMOMC; Rachel 
Bean and Olivier Dor\'e  for useful discussions, and the anonymous referee 
for useful comments.

\end{document}